\documentclass[a4paper,12pt]{article}
\pdfoutput=1 

\usepackage{jheppub} 

\usepackage[T1]{fontenc} 

\title{\boldmath Zero Temperature Dissipation and Holography}

\author{Pinaki Banerjee}
\author{and B. Sathiapalan}
\affiliation{Institute of Mathematical Sciences\\ CIT Campus, Taramani \\ Chennai - 600 113, India}

\emailAdd{pinakib@imsc.res.in}
\emailAdd{bala@imsc.res.in}

\abstract{We use holographic techniques to study the zero-temperature limit of dissipation for a Brownian particle moving 
in a strongly coupled CFT at finite temperature in various space-time dimensions. The dissipative term in the
boundary theory for $\omega\to 0$, $T \to 0$ with $\omega / T$ held small and fixed, does not match the same at 
$T=0$, $\omega \to 0$. Thus the $T\to 0$ limit is not smooth for $\omega < T$. This phenomenon
appears to be related to a confinement-deconfinement phase transition
at $T=0$ in the field theory.}

\begin{document} 
\maketitle
\flushbottom

\section{\label{sec:intro}Introduction}
The motion of an external heavy quark in a non-Abelian
gauge theory has been studied 
in a number of papers (see \cite{Mukund,Son-Teaney,Us,PB, Pedraza3,Hatta,Chernicoff3,Chernicoff4,
Chernicoff1,Chernicoff2,Chernicoff,HKKKY,Pedraza2} and references there in). One motivation for this
are the experimental results that came out of the Relativistic Heavy Ion Collider (RHIC).
The suggestion that the quark gluon plasma
is ``strongly coupled'' with a very small value of $\eta\over s$ came from experiments. 
The calculation of $\eta\over s$ using AdS/CFT techniques \cite{Maldacena,Witten1,GKP,Witten2} gave the small value of
$1\over 4\pi$ \cite{PSS}. This gave impetus to holographic techniques
for understanding
the quark gluon plasma. On the dual gauge theory side  exact calculations have been done 
for $\mathcal {N}$=4 super Yang-Mills theory - calculations that make heavy use of supersymmetry. One 
can hope that at finite temperature the deconfined QCD quark gluon plasma behaves qualitatively 
like strongly coupled $\mathcal {N}$=4 super Yang-Mills theory. The ``running'' effective coupling constant
of QCD {\em presumably} is large in the field configurations that dominate the quark gluon plasma and
therefore well approximated by a {\em strongly} coupled super Yang-Mills theory. At finite 
temperature since supersymmetry is broken anyway and fermions and scalars effectively become 
massive, one can also presume that supersymmetry and the presence of adjoint fermions and scalars, 
does not invalidate the approximation. The success of the holographic calculation provides 
some indirect justification for all this. \\

Many  calculations in super Yang-Mills have been done  at zero temperature where 
supersymmetry is exact. In flat space, at all non zero temperatures, the theory is in the 
same Coulomb phase (unlike QCD) and therefore calculations done at zero temperature
might be thought to still be of some relevance for the quark gluon plasma.
On the dual gravity side this corresponds
to pure AdS$_5$. On the gravity side it is a little easier to explore the finite temperature regime \cite{Witten2}.
This corresponds to non-extremal D3-branes. In fact the $\eta\over s$ calculation involves 
calculating $\eta$ and $s$ separately at finite temperature. \\

This paper studies the $T\to 0$ limit of the finite temperature calculations on the gravity side.
This limit is subtle for reasons that will become clear later. The motivation for this study 
comes from an earlier paper \cite{Us} where the Langevin equation describing Brownian 
motion\footnote{Brownian motion of a heavy quark in quark-gluon-plasma was first described using
holographic techniques in \cite{Mukund,Son-Teaney}.} of a stationary heavy quark in 1+1 CFT at finite temperature
was studied using the gravity dual, which is a BTZ black hole. The calculation was done using 
the holographic Schwinger-Keldysh method  
worked out in\cite{ Son-Teaney}. The calculation can be done exactly (unlike in 3+1 dimensions).
One of the interesting results is that there is a drag force (dissipating energy) on the
fluctuating external quark {\em even at zero temperature}. This was identified as being due to radiation\cite{Mikhailov,Xiao,
Pedraza3,Hatta,Chernicoff3,Chernicoff4,Chernicoff1,Chernicoff2,Chernicoff,Fulton,Boulware,Correa}.
This force term in the Langevin equation  was of the form
\[
F(\omega) =-i {\sqrt{\lambda}\over 2\pi}\omega^3 x(\omega)
\]
If one calculates the integrated energy loss one finds ($x(t)=\int {d\omega\over 2\pi}x(\omega)e^{-i\omega t}$):
\[
\Delta E = \int _{-\infty}^\infty F(t) \dot x(t) dt = \int {d\omega\over 2\pi} F(w)(i \omega) x(-\omega)=\int {d\omega\over 2\pi} (-i) {\sqrt{\lambda}\over 2\pi}\omega^3 x(\omega)(i \omega) x(-\omega)
\]
\[
={\sqrt{\lambda}\over 2\pi}\int {d\omega \over 2\pi} \omega^4 x(\omega)x(-\omega)
\]

While the above calculation was done by taking the $T\to 0$ of a finite temperature calculation 
in BTZ, the same result is obtained for pure AdS in all dimensions, as shown in this paper. 
The energy radiated by an accelerating quark has been calculated using other techniques 
(also holographic) by Mikhailov \cite{Mikhailov} and the answer obtained is
\[
\Delta E = {\sqrt{\lambda}\over 2\pi}\int dt~a^2
\]
which on Fourier transforming gives exactly the same result. In fact the coefficient 
${\sqrt{\lambda}\over 2\pi}$ is essentially the bremsstrahlung function\footnote{see \cite{Fiol0,Fiol1,Fiol2,Fiol3} and references 
there in for more details about bremsstrahlung function in supersymmetric theories.} 
$B(\lambda,N)$ ($2\pi B(\lambda,N)={\sqrt{\lambda}\over 2\pi} $) identified 
in \cite{LM} as
occurring in many other physical quantities (such as the cusp 
anomalous dimension introduced by Polyakov\cite{Polyakov}). \\

It is thus interesting to check whether the same result is also obtained as one takes $T\to$ 0
in 3+1 dimensions. In fact the $T\to 0$ limit is a little tricky because of singularities. 
Taking $T \to 0$ in the finite $T$ theory can be done if one keeps $\omega \over T$ small. Thus the limit where 
$\omega \to 0$ and $T \to 0$ is well defined and can be calculated perturbatively. (We shall see later that all the 
dimensionful quantities are measured in the unit of quark mass, $\mu$). The result for dissipation in this (DC)  limit can be compared with
that calculated for the $T=0$ result of pure AdS for $\omega \to 0$. The results do not agree. Thus we find that
pure AdS ($T=0$) results cannot be assumed to be close to small $T$ results as one would naively have assumed. 
This is not surprising given that there is a Hawking Page
transition\cite{Hawking-Page} at exactly zero temperature in the Poincare patch description of Schwarzschild-AdS.
(In global AdS where space is $S^3$, this happens at a finite temperature.) But this does raise 
questions of the relevance of the zero temperature calculations in $\mathcal{N}$=4 super Yang-Mills for 
comparison with data taken from experiments such as RHIC. \\

In this paper we study in a general way, the zero temperature limit of some theories in 3+1
dimensions using holography. We show how one can do a perturbation in $T$ - however this has to 
be done {\em about a solution that is singular as $T\to 0$}. This does not go smoothly to 
the $T=0$ result of pure AdS. \\

The rest of the paper is organized as follows. In section \ref{sec:zero} we discuss dissipation at exactly zero temperature
by studying dynamics of a long fundamental string in pure AdS space-time. We check whether the dissipation very close to zero
temperature smoothly matches the same at absolute zero in section \ref{sec:near-zero} by changing the background geometry
for the moving string to AdS-black holes. In section \ref{sec:disc} we try to interpret our results. The section \ref{sec:conc}
summarizes the main results obtained in this paper. The perturbative technique used in AdS$_5$-BH case is also applied 
for studying string in BTZ background as a check of applicability of the method in appendix \ref{sec:A}.

\section{\label {sec:zero}Dissipation at Zero Temperature ($T=0$) }

A Brownian particle dissipates energy at zero temperature only by radiating soft or massless modes (photons, gluons etc).
The dual background where the string moves is a pure AdS space.
\begin{align}\label{AdS metric}
ds^2=- \frac{r^2}{L^2} \mathrm d t^2+ \frac{L^2}{r^2} \mathrm d r^2+ \frac{r^2}{L^2} \mathrm d \vec{x}^2
\end{align} 
$L$ is AdS radius and $\vec{x} \equiv (‎x^1,x^2 \ldots x^{d-1})$. \\

A stochastic string 
in this pure AdS background is exactly solvable in arbitrary dimensions. To compute the retarded Green's function and 
the dissipative term from that we need to study string dynamics in \eqref{AdS metric}.\\

We shall be eventually working with linearized Nambu-Goto action. Therefore after choosing the static gauge,
without loss of generality, we can pick up any one transverse direction
($x^1\equiv x$, say) and fix others ($x^2,x^3 \ldots x^{d-1}$) to zero. Essentially we are looking at a three 
dimensional slice of AdS$_{d+1}$. So, X($\sigma,\tau$) is a map to ($\tau,r,x$). The Nambu-Goto action for small
fluctuation in space and in time reduces to

\begin{align}
S &= - \frac{1}{2\pi l_s^2} \int \mathrm d t \mathrm d r \sqrt{1+\dot{x}^2+\frac{r^4}{L^4} {x^\prime}^2} \nonumber\\
&\approx - \int \mathrm d t \mathrm d r \left[1+ \frac{m}{2}\dot{x}^2+\frac{1}{2} T_0(r) {x^\prime}^2 \right]
\end{align}
 
$m= \frac{1}{2\pi l_s^2}$ and $ T_0(r)= \frac{r^4}{2\pi l_s^2 L^4}$ . Varying the action we get the EOM which 
in frequency space reads
\begin{align}
f_\omega^{\prime \prime}(r)+\frac{4}{r} f_\omega^{\prime}(r) + \frac{L^4 \omega^2}{r^4} f_\omega(r)=0
\end{align}
where $x(r,t)= \int \frac{\mathrm d \omega}{2 \pi } e^{-\, i \omega t}f_\omega(r) x_0(\omega)$ and $x_0(\omega)$ is the 
boundary value\footnote{Notice that $r=r_B$ is the boundary of the geometry. This is IR cutoff for the bulk and 
UV cutoff for the dual field theory. For large but finite value of $r_B$ the quark is very heavy but has a finite mass,
and therefore, a detectable Brownian motion} of $x(r,t)$ such that $f_\omega(r_B) = 1$. \\

This is a linear second order ordinary differential equation with following two linearly independent solutions
\begin{align}
f_\omega^{(1)}(r)= \frac{e^{-i\frac{L^2\omega}{r}}(r+ i L^2\omega)}{r} \\
f_\omega^{(2)}(r)= \frac{e^{i \frac{L^2\omega}{r}}(r- i L^2\omega)}{r} 
\end{align}
\begin{itemize}
\item
As we want to compute \emph{retarded} Green's function we pick $f_\omega^{(2)}(r)$ which is 
\emph{ingoing}\footnote {Notice that $e^{-i\omega t} e^{i\frac{L^2\omega}{r}}=e^{-i\omega(t-\frac{L^2}{r})}$.
To keep the phase unchanged, for increasing $t, ~r$ must decrease. So the wave is ingoing.} at $r=0$. \\

\item
The boundary condition, $f_\omega(r) \to 1$ as $r \to r_B $, fixes the solution to be 
\begin{align} 
f_\omega(r)= \frac{r_B}{r}~ \frac{e^{+\,i\frac{L^2\omega}{r}}(r-iL^2\omega)}{e^{+\,i\frac{L^2\omega}{r_B}}(r_B-iL^2\omega)}
\end{align}
\end{itemize}

We just use these modes in calculating the on-shell action and obtain 
the retarded Green's function\cite{SS,HS}. 
\begin{align}
G^0_R(\omega) &\equiv \lim_{r \to r_B } T_0(r) f_{-\omega}(r) \partial_r f_{\omega}(r) \nonumber\\
&=- \frac{r_B^2 \omega^2}{2\pi l_s^2} \frac{1}{(r_B - iL^2 \omega)} \nonumber\\
&=- i\frac{\mu^2 \omega^2}{2\pi \sqrt{\lambda}} \frac{1}{(\omega + i\frac{\mu}{\sqrt{\lambda}})}
\end{align} 
where we have introduced a mass scale $\mu=\frac{r_B}{l_s^2}$ 
and a dimensionless parameter $\sqrt{\lambda}=\frac{L^2}{l_s^2}$. Notice that $\mu$ and $\lambda$ behave like 
Wilsonian cutoff scale and dimensionless coupling for the field theory respectively \cite{Us}. $\mu$ is 
essentially the energy stored in the string as it is stretched from the horizon to the boundary of the 
AdS. This is also interpreted as the mass of the external quark. \\

For $\mu \to \infty$ , $G^0_R(\omega) = -\frac{\mu \omega^2}{2\pi}$ which is divergent. 
We can renormalize the Green's function by absorbing the UV divergent piece to define the zero 
temperature mass of the quark, i.e, $M^0_Q= \frac{\mu}{2\pi}$ and obtain the \emph{renormalized}
Green's function 
\begin{align} \label{AdS GF}
G_R(\omega)\equiv G^0_R(\omega) +\frac{\mu \omega^2}{2\pi} = \frac{\mu \omega^3}{2\pi} \frac{1}{(\omega + i\frac{\mu}{\sqrt{\lambda}})}
\end{align} 

The renormalized Green's function is UV finite by construction. If we take $\mu \to \infty$ 
(or we can take $\omega$ very small)
\begin{align}
G_R(\omega) \to -i \frac{\sqrt{\lambda}}{2 \pi} \omega^3
\end{align}

This is purely dissipative term which is independent of temperature. 

\section{\label {sec:near-zero}Dissipation near Zero Temperature ($T \to 0$) }

To describe a Brownian particle moving in a d-dimensional space-time at finite temperature holographically one needs
to consider a fundamental string in (d+1)-dimensional dual geometry with a black hole. In this section 
also we work in the Poincare patch of AdS$_{d+1}$-Black hole geometry. \\

We start with the AdS$_{d+1}$-black brane metric \cite{HKKKY}
\begin{align} \label{black-brane}
ds_{d+1}^2= L^2\left[- h(u) d t^2 + \frac{\mathrm d u^2}{h(u)} + u^2 \mathrm d {x}^2\right]
\end{align}
where $h(u)=u^2 (1-\left(\frac{u_h}{u}\right)^d)$ with $u_h=\frac{4 \pi T}{d}$. $u$ has dimension of 
energy. \\

We choose $d=2$ and $4$ for 
illustration. The aim is to
check whether the dissipative terms match smoothly to that of the zero temperature case as we take
$T \to 0$.

\subsection{\label{sec:BTZ }BTZ Black hole Background}
A string in (2+1) dimension is \emph{exactly} solvable even in presence of a (BTZ) black hole \cite{Mukund,Us}. We 
work in \eqref{black-brane} for $d=2$ but with $r$-coordinate where $r \equiv L^2 u$ has dimension of length. 

\begin{align} \label{BTZ}
ds^2=-\frac{r^2}{L^2}\left(1- \frac{(2 \pi T L^2)^2}{r^2}\right) \mathrm d t^2+ \frac{L^2}{r^2} \frac{\mathrm d r^2}{\left(1- \frac{(2 \pi T L^2)^2}{r^2}\right)}+ \frac{r^2}{L^2} \mathrm d \vec{x}^2
\end{align}

Using the holographic prescription for Minkowski space \cite{SS,Son-Teaney} one  obtains the exact 
retarded Green's function. Here are the key steps (see \cite{Us} for details). \\

Choosing the static gauge, we study small fluctuations of the string from the Nambu-Goto action 
\begin{align} \label{NG}
S \approx -\int \mathrm d t \mathrm d r \left [m + \frac{1}{2}T_0 {(\partial_r x)}^2 -\frac{m}{1-\left(\frac{2 \pi T L^2}{r}\right)^2}{(\partial_t x)}^2 \right ]
\end{align}
where, $m \equiv \frac{1}{2 \pi l_s^2 } $ and $T_0(r)\equiv \frac{1}{2 \pi l_s^2 } \frac{r^4}{L^4}\left[1-\left(\frac{2 \pi T L^2}{r}\right)^2\right]$ \\

Now varying this action one obtains the EOM in frequency space 
\begin{align}\label{EOM BTZ}
f_\omega^{\prime \prime}(r) + \frac{2(2r^2- 4 \pi^2 T^2 L^4)}{r(r^2-4 \pi^2 T^2 L^4)}f_\omega^{\prime}(r) + \frac{L^4 \omega^2}{(r^2-4 \pi^2 T^2 L^4)^2} f_\omega(r) =0 
\end{align}
which is exactly solvable and the solution is $$f_\omega(r)= C_1 \frac{P^{\frac{i \omega}{2 \pi T}}_1(\frac{r}{2\pi T L^2})}{r} + C_2 \frac{Q^{\frac{i \omega}{2 \pi T}}_1(\frac{r}{2\pi T L^2})}{r} $$
where $P$ and $Q$ are associated Legendre functions. \\

Now choosing ingoing boundary condition (to obtain retarded Green's function) at the horizon 
and fixing $f_\omega(r_B)=1$ one obtains the required solution   
\begin{align}
f^R_\omega(r)= \frac{(1-\frac{r}{2\pi T L^2})^{-i\omega/4 \pi T}}{(1-\frac{r_B}{2\pi T L^2})^{-i\omega/4 \pi T}}  \frac{(1+\frac{r}{2\pi T L^2})^{i\omega/4 \pi T}}{(1+\frac{r_B}{2\pi T L^2})^{i\omega/4 \pi T}} \frac{r_B}{r} \frac{(L^2 \omega +i r)}{(L^2 \omega + i r_B) }
\end{align} 
Now from the on-shell action we can read off the Green's function
\begin{align} \label {RGF}
G^0_R \equiv \lim_{r \to r_B } T_0(r)f^R_{-\omega}(r) \partial_r f^R_{\omega}(r) = -\mu \omega ~{(i \sqrt \lambda~ 4 \pi ^2 T^2 + \mu \omega )\over 2\pi  ( \mu - i \sqrt \lambda \omega )} 
\end{align} 
where we have used previously defined mass scale $\mu$ and the dimensionless parameter $\lambda$.
For $\mu \to \infty$, $G^0_R(\omega) = -\frac{\mu \omega^2}{2\pi}$ which is again divergent. 
We can renormalize the Green's function as before by absorbing the UV divergent piece to define the zero 
temperature mass of the quark, i.e, $M^0_Q= \frac{\mu}{2\pi}$ and obtain the renormalized
Green's function

\begin{align}\label{BTZ GF}
G_R(\omega)&\equiv G^0_R(\omega) +\frac{\mu \omega^2}{2\pi}= {\mu \omega \over 2\pi}~{(\omega^2 + 4 \pi ^2 T^2)\over (\omega + i {\mu\over \sqrt \lambda})}  
\end{align} 

Near zero temperature
\begin{align}\label{BTZ GF 0}
G_R(\omega) \bigg{|}_{T \to 0} =~ \frac{\mu \omega^3}{2\pi} \frac{1}{(\omega + i\frac{\mu}{\sqrt{\lambda}})}
\end{align} 
This is identical to the retarded Green's function for zero temperature system \eqref{AdS GF}. 
Evidently  for small frequency $$ G_R(\omega) \approx - i\frac{\sqrt{\lambda}}{2 \pi} ~\omega^3 $$

From this calculation black hole background seems to match smoothly pure AdS space as one takes $T \to 0$. 
The obvious question comes to one's mind is whether this coefficient of zero temperature dissipation is universal and independent
of the dimensionality of space time. Actually we will see in the next section that this is not really the case in general. 
Reason being the Poincare patch of BTZ 
black hole is, strictly speaking, AdS$_3$ at finite temperature for all practical purposes. That's why it smoothly matches 
the pure AdS result as one takes $T \to 0$. But AdS$_{d+1}$-BH with 
$d>2$ is a ``genuine'' black hole background and therefore the limit might not be smooth.

\subsection{\label{sec:AdS5-BH}AdS$_5$ Black Hole Background}

Now let us check if the dissipation coefficients for higher dimensional black holes in AdS space near
zero temperature match the exact zero temperature coefficient. But those are not exactly solvable.
As an example we will demonstrate it for AdS$_5$ black hole \cite{Son-Teaney}. \\

We can write down the metric in $r$-coordinate, as before, fixing $d=4$ in \eqref{black-brane}
\begin{align} \label{AdS5 BH}
ds^2=-\frac{r^2}{L^2}\left(1- \frac{(\pi T L^2)^4}{r^4}\right) \mathrm d t^2+ \frac{L^2}{r^2} \frac{\mathrm d r^2}{\left(1- \frac{(\pi T L^2)^4}{r^4}\right)}+ \frac{r^2}{L^2} \mathrm d \vec{x}^2
\end{align}

with $\vec{x}\equiv (x^1,x^2,x^3)$. \\

The EOM\footnote{Notice that for this case $m \equiv \frac{1}{2 \pi l_s^2 } $ and $T_0(r)\equiv \frac{1}{2 \pi l_s^2 } \frac{r^4}{L^4}\left[1-\left(\frac{\pi T L^2}{r}\right)^4\right]$. 
But eventually we want to calculate Green's function for $T \to 0$. Therefore for both BTZ and AdS-BH 
background we can practically use $T_0(r)\equiv \frac{1}{2 \pi l_s^2 } \frac{r^4}{L^4}$.} for the string 
\begin{align} \label{EOM AdS5 BH}
F_\omega^{\prime \prime} (r) + \frac{4r^3}{(r^4-\pi^4 T^4 L^8)}F_\omega^{\prime}(r) + \frac{\omega^2 L^4 r^4}{(r^4-\pi^4 T^4 L^8)^2} F_\omega(r) =0 
\end{align}
can not be solved exactly. As it's an ordinary second order linear differential equation it has 
two linearly independent solutions. \\

Near the horizon $$ F_\omega(r) \sim \left(1-\frac{\pi^4 T^4 L^8}{r^4} \right)^{\pm i \frac{\Omega}{4}}$$ 
\par 
Near the boundary 
\begin{align} \label{near bdy1}
F_\omega(r)=\left(1+\frac{\Omega^2}{2r^2} + \ldots \right) + \frac{\chi(\Omega)}{r^3}\left(1-\frac{\Omega^2}{10r^2} + \ldots \right)
\end{align}
contains a non-normalizable and a normalizable mode. Here $\Omega \equiv \frac{\omega}{\pi T}$. \\

But still one can solve it in perturbation expansion\footnote{The -ve sign in the exponent is chosen in \eqref{ansatz}
which is ingoing at the horizon. Because we are interested to calculate the retarded Green's function.} in small frequency

\begin{align} \label{ansatz}
F_\omega(r)= \left(1-\frac{\pi^4 T^4 L^8}{r^4} \right)^{- i \frac{\Omega}{4}}(1 - i \Omega f_1(r)-\Omega^2 f_2(r) + i \Omega^3 f_3(r)+ \ldots) 
\end{align} 

Putting this ansatz into \eqref{EOM AdS5 BH} we get hierarchy of differential equations.
Solving them order by order in $\Omega \equiv \frac{\omega}{\pi T}$ in principle one
can obtain the unknown functions, $f_i (r)$ where $i = 1, 2, 3, \ldots $ \\

Few useful remarks on the perturbative solution before we actually obtain it. \\

\begin{enumerate}

 \item This type of perturbative solution has been calculated in \cite{Son-Teaney} by Son and Teaney. 
 As the authors were mostly interested in finite temperature phenomena they computed the Green's function 
 up to $\omega^2$ term. In this article we show that their solution can be used even for $T \to 0$ 
 and also we compute the Green's function up to  $\omega^3$ term which indicates the zero temperature 
 dissipation. \\
 
 \item The solution we obtain is a perturbation in $\Omega$ and $T$. But finally we are interested in
 $T \to 0$ limit. Clearly this limit is pathological for any finite $\omega$. Only way we can make 
 sense of the solution is by taking both 
 $$ \omega, T \to 0~  \text{~with~} ~\Omega ~\text{held fixed (and small).} $$ 
 The temperature independent term that we 
 are interested in is the coefficient of $\Omega^3 T^3$ in this solution. \\
 
 It is important to note, all dimensionful quantifies in this theory are measured in terms of 
 the cutoff scale $\mu ~( = \frac{r_B}{l_s^2})$ defined before which is also interpreted as 
 the mass of the external quark. Therefore whenever we say
 $\omega, T \to 0~  \text{~with~} \Omega ~\text{held fixed}$ we mean 
 $\omega/\mu \to 0$ and $T/\mu \to 0$ such a way that $\omega/T$ is fixed small number. E.g, say, 
 $\omega/\mu =10^{-7},~~ T/\mu =10^{-6}$, thus $\omega/T = 0.1$ which is smaller than 1.\\
 
 \item Actually, as we will see below, we don't need to obtain all the unknown functions, 
 $f_i (r)$, explicitly by performing complicated integrals. Rather we need only the \emph{residues} of those 
 integrals at the horizon to fix the coefficient of the homogeneous solutions. \\

\end{enumerate} 
\subsubsection*{Perturbative solution in AdS$_5$-BH}
Just for convenience we work with $z$ co-ordinate, where $z \equiv \frac{L^2}{r}$. Obtaining results in the 
$r$ variable is straightforward. 
As we have discussed the EOM \eqref{AdS5 BH} for the string in AdS$_5$-BH
\begin{align} \label{A:EOM}
F_\omega^{\prime \prime} (z) + \frac{2 (1 + \pi^4 T^4 z^4)}{z(1-\pi^4 T^4 z^4)}F_\omega^{\prime}(z) + \frac{\omega^2}{(1-\pi^4 T^4 z^4)^2} F_\omega(z) =0 
\end{align}
is not exactly solvable and to obtain the solution that is ingoing at the horizon we need to use the 
following ansatz.
\begin{align} \label{A:ansatz}
F^R_\omega(z)&= \left(1- \pi^4 T^4 z^4 \right)^{- i \frac{\Omega}{4}} H(z) \nonumber \\
\text{where~~} H(z) &= 1 - i \Omega h_1(z)-\Omega^2 h_2(z) + i \Omega^3 h_3(z)+ \ldots
\end{align}
with $\Omega \equiv \frac{\omega}{\pi T}$. \\

The differential equation $H(z)$ satisfies is given by

\begin{align}\label{A:DE-H}
H^{\prime \prime} (z) - \frac{2 \left(1+\pi ^3 T^3 z^4 (\pi  T-i \omega )\right)}{z \left(1 - \pi^4\, T^4\, z^4 \right)} H^{\prime}(z) + \frac{\omega  \left(\omega +\pi ^2 T^2 z^2 (\omega +i \pi  T) \left(\pi ^2 T^2 z^2+1\right)\right)}{\left(\pi ^2 T^2 z^2+1\right) \left(1- \pi ^4 T^4 z^4\right)} H(z) = 0
\end{align}

Notice that by choosing the ansatz \eqref{A:ansatz} we have taken care of the singular near horizon part of the full solution
by the pre-factor $\left(1- \pi^4 T^4 z^4 \right)^{- i \frac{\Omega}{4}}$.
Our strategy would be to substitute the ansatz \eqref{A:ansatz} into \eqref{A:DE-H} and at each
order in $\Omega$ we demand that the solution to \eqref{A:DE-H}  is \emph{regular} at the \emph{horizon}. In other words, 
the full solution to \eqref{A:EOM} at any order in $\Omega$ behaves 
like $$\left(1-\pi^4 T^4 z^4 \right)^{- i \frac{\Omega}{4}} \times \left(\text{Regular
function at } z=\frac{1}{\pi T}\right)$$

Again, we are interested in calculating temperature independent dissipative term in the Green's function.
Therefore on dimensional ground we need to look for the coefficient of $\omega^3$ term in the Green's function
as the Green's function itself has mass dimension three. In other words, if one takes zero temperature limit
of the Green's function only $\omega^3$ term survives. For that one needs to take $T \to 0$ limit of the solution. 
But clearly the solution is a perturbation series in $\Omega = \frac{\omega}{\pi T}$. Therefore the only way 
one can make sense of this solution near zero temperature is to take both $T \to 0$ and $\omega \to 0$
(compared to $\mu$) keeping the perturbation parameter, $\Omega$ fixed and small ($\Omega < 1$) such that the series converges. \\

\begin{figure}[h]
    \centering
    \includegraphics[width=8cm]{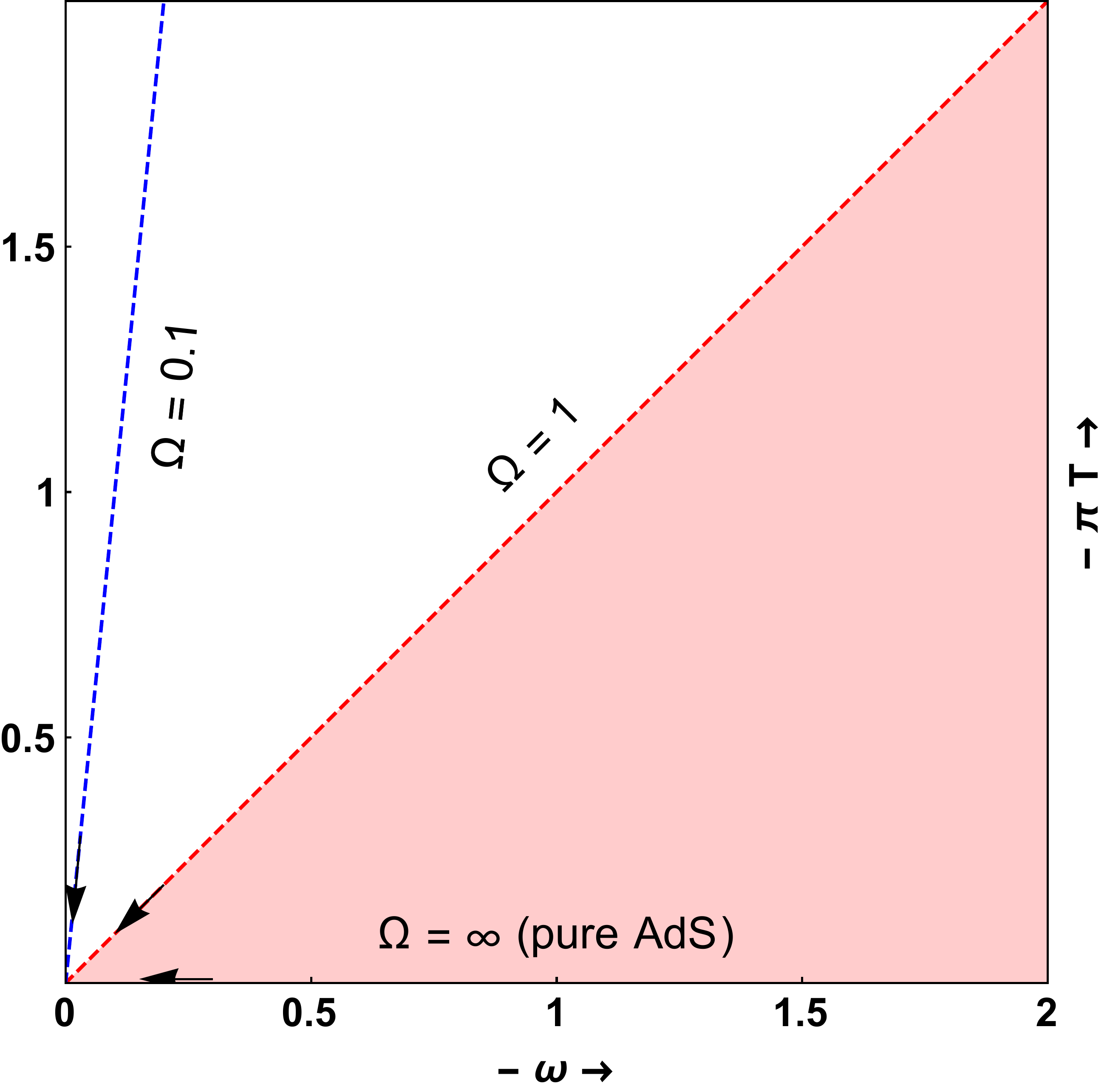}
    \caption{Different ways of taking $T=0, \omega=0$ limit.}
    \label{fig:W-T}
\end{figure}

Pictorially, there are many ways one can reach $(\omega=0, T=0)$ i.e, the origin on the $\omega$-$
T$ plane (see figure \ref{fig:W-T}). Our solution makes
sense when $\omega/\pi T$ is a constant and smaller than one. The shaded region is outside the domain of validity
of our solution. Thus all our analysis and results hold true for the straight lines in the upper half of the box. \\

Also notice that pure AdS ($T=0$) is along the $x$-axis. Therefore eventually we will be comparing these two
ways ($\Omega$ = small and $\Omega = \infty$) of taking limits. \\

\textbf{\underline{Solution up to $\mathcal{O}(\Omega)$} :}
\begin{align}
 H(z) = 1 - i \Omega h_1(z)
\end{align}

Substituting this ansatz into \eqref{A:EOM} we obtain the differential equation for $h_1(z)$ 
\begin{align}
h_1^{\prime \prime} (z) + \underbrace{\frac{2 (1 + \pi^4 T^4 z^4)}{z(1-\pi^4 T^4 z^4)}}_{p_1(z)}h_1^{\prime}(z) = \underbrace{\frac{\pi^4 T^4 z^2}{(1-\pi^4 T^4 z^4)}}_{q_1(z)}
\end{align}

Lets cast this into a first order differential equation defining  $y_1(z) \equiv h_1^{\prime } (z)$ and consequently $y_1^{\prime}(z) \equiv h_1^{\prime \prime} (z)$

\begin{align}
y_1^{\prime} (z) + {p_1(z)} y_1(z) = {q_1(z)}
\end{align}

One can introduce integrating factor $I_1(z) = \exp{(\int p_1(z) \mathrm d z)}$ to obtain
\begin{align}
y_1(z)&= \frac{c_1}{I_1(z)} +  \frac{1}{I_1(z)} \int^z I_1(x) q_1(x) \mathrm d x \nonumber \\
&= \underbrace{\frac{c_1 z^2}{1-\pi^4 T^4 z^4}}_{y_1^h(z)} + \underbrace{\frac{\pi^4 T^4 z^3}{1-\pi^4 T^4 z^4}}_{y_1^p(z)}
\end{align}

We will see that the homogeneous part of the solution ($y_i^h(z)$) is identical in each order
in $\Omega$ up to the undetermined coefficient ($c_i$). This coefficient is fixed by demanding
the regularity of $f_i$ at the horizon. 
\begin{align}
 f_1(z)&=  \int y_1^h(z)\mathrm d z + \int  y_1^p(z)\mathrm d z \\
&\equiv h_1^h(z) + h_1^p(z)
\end{align}

The requirement that $h_1(z)$ has to be regular sets the coefficient of $Log(1-\pi T z)$ to zero. 
One can explicitly calculate the integrals and from that expression sort out the required coefficient.
For this case 
\begin{align}
 h_1^h(z) &= \frac{c_1}{4 \pi^3 T^3} \left\{-2 tan^{-1}(\pi T z) - Log (1-\pi T z)+ Log(1+\pi T z)\right\} \\
 h_1^p(z) &= -\frac{1}{4} Log(1-\pi^4 T^4 z^4) \nonumber \\
 &= -\frac{1}{4} \left\{Log (1-\pi T z)+ Log(1+\pi T z)+ Log(1+\pi^2 T^2 z^2)\right\}
\end{align}

Clearly setting the coefficient of $Log (1-\pi T z)$ to zero we get $$c_1 = -\pi^3 T^3$$ 
And the solution at this order becomes
\begin{align}\label{A:h1}
  h_1(z) = \frac{1}{2} tan^{-1}(\pi T z) - \frac{1}{2} Log(1+\pi T z) + \frac{1}{4} Log(1+\pi^2 T^2 z^2)
\end{align}

But there is another way in which we can fix the coefficient without doing the integrals.
The only potential singular term in $h_1(z)$ at the horizon appears as $Log (1-\pi T z)$. 
This type of terms are originated from the terms of the form $\frac{1}{(1-\pi T z)}$ in $y_1(z)$. Therefore 
fixing the coefficient of $Log (1-\pi T z)$ in $f_1(z)$ to zero is equivalent to setting
the residue of $y_1(z)$ at $z=\frac{1}{\pi T}$ to zero. This is much easier way when the
integrals get complicated as we go in higher orders. \\

\textbf{\underline{Solution up to $\mathcal{O}(\Omega^2)$} :}

\begin{align}
H(z)= 1 - i \Omega h_1(z) -\Omega^2 h_2(z)
\end{align}
Notice that $h_1(z)$ is already known from \eqref{A:h1}. Substituting this ansatz into \eqref{A:DE-H}
we get the differential equation for $h_2(z)$. Again one can cast that into a first order differential equation

\begin{align}
y_2^{\prime} (z) + {p_2(z)} y_2(z) = {q_2(z)}
\end{align}
where $p_2(z)=p_1(z)$. Therefore integrating factor $I_2(z)=I_1(z)$. 

\begin{align}
y_2(z)&= \frac{c_2}{I_2(z)} +  \frac{1}{I_2(z)} \int^z I_2(x) q_2(x) \mathrm d x \nonumber \\
&= \underbrace{\frac{c_2 z^2}{1-\pi^4 T^4 z^4}}_{y_2^h(z)} + \underbrace{\frac{1}{I_2(z)} \int^z I_2(x) q_2(x) \mathrm d x}_{y_2^p(z)}
\end{align}

Now making the residue of $y_2(z)$ at $z=\frac{1}{\pi T}$ to vanish we can fix  $$c_2 = \pi^3 T^3$$
The solution at this order

\begin{align}\label{A:h2}
  h_2(z) = \frac{1}{32}[&4\{-4 + tan^{-1}(\pi T z)-Log(1+\pi T z)\}\{tan^{-1}(\pi T z)-Log(1+\pi T z) \} \nonumber \\
  &- 4 \{ 2 + tan^{-1}(\pi T z) - Log(1+\pi T z)\}Log(1+\pi^2 T^2 z^2)+Log(1+\pi^2 T^2 z^2)^2]
\end{align} \\

\textbf{\underline{Solution up to $\mathcal{O}(\Omega^3)$} : }

\begin{align}
 H(z) = 1 - i \Omega h_1(z)-\Omega^2 h_2(z) + i \Omega^3 h_3(z)
\end{align}
where $h_1, h_2$ are known from \eqref{A:h1} and \eqref{A:h2}. The differential equation for
$h_3(z)$ or rather $y_3(z) \equiv h_3^{\prime}(z)$
\begin{align}
y_3^{\prime} (z) + {p_3(z)} y_2(z) = {q_3(z)}
\end{align}
where $p_3(z)=p_1(z)$. Therefore integrating factor $I_3(z)=I_1(z)$, as before. 

\begin{align}
y_3(z)&= \frac{c_3}{I_3(z)} +  \frac{1}{I_3(z)} \int^z I_3(x) q_3(x) \mathrm d x \nonumber \\
&= \underbrace{\frac{c_3 z^2}{1-\pi^4 T^4 z^4}}_{y_3^h(z)} + \underbrace{\frac{1}{I_3(z)} \int^z I_3(x) q_3(x) \mathrm d x}_{y_3^p(z)}
\end{align}

Now making the residue of $y_3(z)$ at $z=\frac{1}{\pi T}$ to vanish we obtain 
$$c_3 =\left(\frac{\pi - Log\, 4}{4}\right) \pi^3 T^3$$ \\

The functional form of $y_3(z)$ is complicated and therefore it's difficult to 
obtain an explicit expression for $h_3(z)$ unlike the lower order functions. But as we are 
interested in zero temperature limit of the Green's function, we only need to know 
the full solution for small $\Omega$

\begin{align}
F^R_\omega(z)&= \left(1- \pi^4 T^4 z^4 \right)^{- i \frac{\Omega}{4}}(1 - i \Omega h_1(z)-\Omega^2 h_2(z) + i \Omega^3 h_3(z))\nonumber \\
&\approx \left(1+\frac{i}{4}\pi^3 \omega T^3 z^4\right)\left\{ 1-\frac{i\omega}{\pi T}(-\frac{1}{3}\pi^3 T^3 z^3)-\frac{\omega^2}{\pi^2 T^2}(-\frac{1}{3}\pi^2 T^2 z^2)-\frac{i \omega^3}{\pi^3 T^3}(\frac{\pi-Log\, 4}{12})\pi^3 z^3 T^3\right\} 
\end{align}

Evidently in the zero temperature limit (with very very small frequency)
\begin{align}\label{A:soln-z}
F^R_\omega(z)\bigg|_{T \to 0} = 1 + \frac{\omega^2 z^2}{2} + i \frac{\omega^3 z^3}{3}\left(\frac{\pi - Log\, 4}{4}\right)
\end{align}

In $r$ co-ordinate 
\begin{align}\label{A:soln-r}
F^R_\omega(r)\bigg|_{T \to 0} = 1 + \frac{\omega^2 L^4}{2 r^2} + i \frac{\omega^3 L^6}{3 r^3}\left(\frac{\pi - Log\, 4}{4}\right)
\end{align}

The retarded Green's function at $T \to 0$ (with $\omega \to 0$ and $\Omega = $ fixed) can be calculated using the solution \eqref{A:soln-r} 
\begin{align}
G^0_R &\equiv \lim_{r \to r_B } T_0(r)F^R_{-\omega}(r) \partial_r F^R_{\omega}(r) \nonumber \\
&=  \lim_{r \to r_B } \frac{1}{2 \pi l_s^2} \frac{r^4}{L^4}\left(-\frac{\omega^2 L^4}{r^3} - i\frac{\omega^3 L^6}{4 r^4}(\pi - Log\, 4)\right)\nonumber\\
&= -\frac{\mu \omega^2}{2 \pi} - i\frac{\sqrt{\lambda}}{2 \pi}\left(\frac{\pi - Log\, 4}{4}\right) ~\omega^3
\end{align} 

Therefore the renormalized Green's function 
\begin{align}
G_R(\omega)\equiv G^0_R + \frac{\mu \omega^2}{2 \pi} =  - i\frac{\sqrt{\lambda}}{2 \pi}\left(\frac{\pi - Log\, 4}{4}\right) ~\omega^3
\end{align}

It is evident that this zero temperature dissipation term is \emph{not} same as that
of pure AdS case and actually off by a factor of $\frac{\pi - Log\, 4}{4}$. 

\section{\label{sec:disc} Discussions}

 Whenever a charged particle accelerates or decelerates it radiates energy which is known as
 bremsstrahlung effect. In this paper we talk about dissipation at and near zero temperature. 
 Naturally this zero temperature dissipation finds its origin in this bremsstrahlung phenomenon. 
 One can notice that this dissipative force term ($\mathcal{F}_{diss}$) goes as a cubic power
 in frequency  $$ \mathcal{F}_{diss}(\omega) \sim - i \sqrt{\lambda} ~\omega^3 x(\omega) $$  
 In real space this each $i \omega$ gives a time derivative and therefore the above force law 
 reduces to $$ \mathcal{F}_{diss}(t) \sim  \sqrt{\lambda}~ \dddot{x}(t) =  \sqrt{\lambda}~ \dot{a}$$
 $\dot{a}$ here quantifies the rate of change in acceleration and is often called jerk or jolt. 
 This formula is very similar to that of ``Abraham-Lorentz force'' \cite{Griffiths} in classical 
 electrodynamics for a charged particle with charge $q$
   $$ \mathcal{F}_{rad}(t) = \frac{2}{3} q^2~ \dot{a}$$
This is the force that an accelerating charged particle feels in the recoil from the emission of radiation.
Only the effective coupling is different for the holographic case. This ``coupling''($\sqrt{\lambda}$) is
essentially the bremsstrahlung function $B(\lambda,N)$. The corrections 
in $\lambda$ and $N$ can also be computed for particular known theories (see \cite{Fiol0,Fiol1,Fiol2,Fiol3}). \\

The main aim in this paper is to understand how this bremsstrahlung function behaves near zero 
temperature. We work in Poincare patch of AdS-black hole on the gravity side. We notice that for higher 
dimensional cases (we performed calculations in AdS$_5$-BH) value of this function\footnote{Of course we are working in 
leading order in large $N$ and large $\lambda$. To compute corrections one needs to work with some known supersymmetric
UV theories (e.g, ABJM, $\mathcal{N}=4$ SYM)} near $T\to 0$ doesn't match that of at $T=0$. Strictly speaking, this result  
is obtained in a particular regime of the parameter space namely  $\omega/\mu \to 0$ and $T/\mu \to 0$
such a way that $\Omega < 1$. Thus the domain of validity of our analysis is spanned by family of straight lines 
(see figure \ref{fig:W-T}) which end 
at the origin and with slope greater than one (i.e, $\frac{\pi \,T}{\omega} > 1$). It is important to note, $\omega \to 0$ limit for 
the solution to the pure AdS case is actually represented by the $x$-axis of figure \ref{fig:W-T}. Therefore effectively we are comparing 
two different ways\footnote{It is worthwhile noting that the lines that end on the origin and are contained inside the shaded 
region are also valid ways of taking $T=0, \omega=0$ limit. But our analysis doesn't work for that region. Because 
in that region $\Omega > 1$ and therefore our perturbative expansion \eqref{A:ansatz} breaks down. Thus we cannot say much about high frequency
domain with this analysis.}  of taking $T=0, \omega=0$ limit, ``almost vertical lines''(AdS-BH) and 
``the horizontal line''(pure AdS). They don't match. Actually this result is not unexpected  as there is a Hawking-Page phase transition 
at zero temperature in Poincare patch of black holes in AdS. \\

Therefore the result suggests that one should not use the $T=0$ theory (which always has $\Omega = \infty$) as an approximation
to small temperature physics, when $\Omega$ is small i.e, $\omega < T$. Thus for example if $T= 10^{-6}$ and $\omega = 10^{-7}$ compared to $\mu$, we cannot
use the $T=0$ theory. This is something that can be very crucial in the context of quark-gluon-plasma (QGP). QGP is always at finite temperature
and therefore dissipation term more specifically the bremsstrahlung function $B(\lambda)$ is not continuously connected to the zero temperature 
background (pure AdS) result, at least for small frequencies ($\Omega < 1$). Therefore one must use the AdS-BH background to compute those quantifies even
at very small temperature. \\

Unlike the higher dimensional case, we see that for a particle in 1+1 dimensional field theory the bremsstrahlung functions match smoothly
at $T=0$. The possible reason behind this phenomenon is hidden in the corresponding dual geometries namely AdS$_3$ (for $T=0$) and
BTZ (at $T\ne 0$). A BTZ black hole is just an orbifold of AdS$_3$ and therefore locally AdS$_3$. 
One can only distinguish the former from the latter by studying 
global properties. BTZ in Poincare patch is no different than AdS$_3$ at finite temperature (also called thermal AdS$_3$) 
unlike the higher dimensional black holes in AdS which are ``genuine'' black hole backgrounds.

\section{\label{sec:conc} Conclusions }

We summarize the main points here: \\

\begin{itemize}
 \item We have studied Brownian motion in various space time dimensions with the help of holographic
       Green's function computation. In each case we obtain dissipation at zero temperature due to 
       radiation from accelerated charged Brownian particle. \\
       
 \item As long as we are considering dynamics of the quark at zero temperature, that is the string in 
       the dual gravity theory moves 
       in a pure AdS spacetime, the coefficients of dissipation for arbitrary space time dimensions 
       are identical. The value of the coefficient is $\frac{\sqrt{\lambda}}{2 \pi}$ and can be identified 
       with $B(\lambda,N)$ \cite{LM} (actually $2\pi B(\lambda,N)={\sqrt{\lambda}\over 2\pi} $) as occurring
       in many other physical quantities (such as the cusp anomalous dimension introduced by Polyakov\cite{Polyakov}). \\
       
 \item Even the coefficients match for string in AdS$_3$ and in BTZ as we take $T\to 0$. This is because BTZ in 
       Poincare patch is nothing more than a thermal AdS$_3$. \\
       
 \item For higher dimensions the coefficients at $T=0$ and $T\to 0$ don't match for small frequencies ($\Omega < 1$).
       Here we are effectively comparing \emph{infinite} $\Omega$  to \emph{finite} and \emph{small} $\Omega$ 
       and they turn out to be different although both refer to the same region around $\omega=0, T=0$. 
       Thus one should be careful in  using pure AdS for calculating near zero temperature quantities (e.g, $B(\lambda)$)
       for very low (near zero) frequencies, i.e. $\Omega < 1$.
       Even if the temperature is very very small (unless it's exactly zero) one should not 
       use $T=0$ results as the $T=0$ and $T \to 0$ systems are described by completely different theories.  
       We have shown this phenomenon via explicit computation by studying a string dynamics in AdS$_5$-BH background. The corresponding coefficient 
       comes out to be  $\frac{\sqrt{\lambda}}{2 \pi} \frac{\pi - Log\, 4}{4}$. This phenomenon might have its origin 
       in the Hawking-Page transition at $T=0$ in Poincare patch.

\end{itemize}


\appendix 

\section{\label{sec:A} Perturbative solution for string in BTZ}

We already mentioned that EOM for a string in BTZ can be solved exactly and hence one can compute 
exact retarded Green's function for the Brownian particle. Actually we have done the same in \ref{BTZ}.
There we have seen that the zero temperature dissipation coefficient is $\frac{\sqrt{\lambda}}{2 \pi}$.
On the other hand, the EOM of string in AdS$_5$-BH is not exactly solvable and therefore we adopted a 
perturbative technique to compute the above mentioned coefficient. In \ref{sec:AdS5-BH} we got a different value  
for that coefficient. In this section we apply the same perturbative method for a string in BTZ and show that
the same result for zero temperature dissipation is reproduced. This is just to show that the perturbative
approach and the associated limits indeed work. \\

Our aim is to perturbatively solve \eqref{EOM BTZ} which in $z$-coordinate looks

\begin{align} \label{B:EOM BTZ}
f_\omega^{\prime \prime} (z) + \frac{2 }{z(1-4 \pi^2 T^2 z^2)}f_\omega^{\prime}(z) + \frac{\omega^2}{(1-4 \pi^2 T^2 z^2)^2} f_\omega(z) =0 
\end{align}
using the following ansatz 
\begin{align} \label{B:ansatz}
f^R_\omega(z)= \left(1- 4\pi^2 T^2 z^2 \right)^{- i \frac{\Omega}{2}}(1 - i \Omega h_1(z)-\Omega^2 h_2(z) + i \Omega^3 h_3(z)+ \ldots) 
\end{align}
where in this section $\Omega \equiv \frac{\omega}{2 \pi T}$ (notice the extra factor of two). \\

Again we will be working in the regime where $$ \omega, T \to 0~  \text{~with~} ~\Omega ~\text{held fixed (and small).} $$ 

Next step is to solve for the unknown functions, $h_1(x), h_2(x), h_3(x)$ etc. recursively. We repeat the same procedure 
as in \eqref{sec:AdS5-BH}. \\

\textbf{\underline{Solution up to $\mathcal{O}(\Omega)$} :}
\begin{align}
 f_\omega(z)= \left(1- 4\pi^2 T^2 z^2 \right)^{- i \frac{\Omega}{2}}(1 - i \Omega h_1(z))
\end{align}

Substituting this ansatz into \eqref{B:EOM BTZ} we obtain the differential equation for $h_1(z)$ 
\begin{align}
h_1^{\prime \prime} (z) - \underbrace{\frac{2}{z(1- 4 \pi^2 T^2 z^2)}}_{p_1(z)} h_1^{\prime}(z) = - \underbrace{\frac{4 \pi^2 T^2}{(1-4 \pi^2 T^2 z^2)}}_{q_1(z)}
\end{align}

Lets cast this into a first order differential equation defining  $y_1(z) \equiv f_1^{\prime } (z)$ and consequently $y_1^{\prime}(z) \equiv f_1^{\prime \prime} (z)$

\begin{align}
y_1^{\prime} (z) + {p_1(z)} y_1(z) = {q_1(z)}
\end{align}

Introducing the integrating factor $I_1(z) = \exp{(\int p_1(z) \mathrm d z)}$ 
\begin{align}
y_1(z)&= \frac{c_1}{I_1(z)} +  \frac{1}{I_1(z)} \int^z I_1(x) q_1(x) \mathrm d x \nonumber \\
&= \underbrace{\frac{c_1 z^2}{1-4 \pi^2 T^2 z^2}}_{y_1^h(z)} + \underbrace{\frac{4 \pi^2 T^2 z}{1-4 \pi^2 T^2 z^2}}_{y_1^p(z)}
\end{align}

The homogeneous part of the solution ($y_i^h(z)$) will again be identical in each order
in $\Omega$ up to the undetermined coefficient ($c_i$). This coefficient is fixed by demanding
the regularity of $h_i$ at the horizon. 
\begin{align}
 h_1(z)&=  \int y_1^h(z)\mathrm d z + \int  y_1^p(z)\mathrm d z \\
&\equiv h_1^h(z) + h_1^p(z)
\end{align}

Demanding regularity for $h_1(z)$ fixes the coefficient of $Log(1- 2 \pi T z)$ to zero. 
One can explicitly calculate the integrals and from that expression sort out the required coefficient.
For this case 
\begin{align}
 h_1^h(z) &= c_1\left[-\frac{z}{4 \pi^2 T^2} + \frac{1}{16 \pi^3 T^3} \left\{Log (1+ 2 \pi T z) - Log(1- 2 \pi T z)\right\} \right] \\
 h_1^p(z) &= -\frac{1}{2} Log(1- 4 \pi^2 T^2 z^2) \nonumber \\
 &= -\frac{1}{2} \left\{Log (1 + 2 \pi T z)+ Log(1 - 2 \pi T z)\right\}
\end{align}

Clearly setting the coefficient of $Log (1- 2 \pi T z)$ to zero we get $$c_1 = -\, 8 \pi^3 T^3$$ 
And the solution at this order becomes
\begin{align}\label{B:h1}
  h_1(z) = \frac{1}{2} \left[ 4 \pi T z - 2 Log(1+ 2 \pi T z) \right]
\end{align}

As has been argued before we can equivalently set the residue of $y_1(z)$ at $z=\frac{1}{2 \pi T}$ to zero 
to fix the value of $c_1$. \\

\textbf{\underline{Solution up to $\mathcal{O}(\Omega^2)$} :}

\begin{align}
 f_\omega(z)= \left(1- 4 \pi^2 T^2 z^2 \right)^{- i \frac{\Omega}{2}}(1 - i \Omega h_1(z) -\Omega^2 h_2(z))
\end{align}
$h_1(z)$ is already known from \eqref{B:h1}. The differential equation for $h_2(z)$ reduces to

\begin{align}
y_2^{\prime} (z) + {p_2(z)} y_2(z) = {q_2(z)}
\end{align}
where $p_2(z)=p_1(z)$ and integrating factor $I_2(z)=I_1(z)$. 

\begin{align}
y_2(z)&= \frac{c_2}{I_2(z)} +  \frac{1}{I_2(z)} \int^z I_2(x) q_2(x) \mathrm d x \nonumber \\
&= \underbrace{\frac{c_2 z^2}{1-4 \pi^2 T^2 z^2}}_{y_2^h(z)} + \underbrace{\frac{1}{I_2(z)} \int^z I_2(x) q_2(x) \mathrm d x}_{y_2^p(z)}
\end{align}

Now making the residue of $y_2(z)$ at $z=\frac{1}{2 \pi T}$ to vanish we can fix  $$c_2 = 8 \pi^3 T^3$$
The solution at this order

\begin{align}\label{B:h2}
  h_2(z) = \frac{1}{2} \left[ -4 \pi T z + \{Log (1 + 2 \pi T z)\}^2 \right]
\end{align} \\

\textbf{\underline{Solution up to $\mathcal{O}(\Omega^3)$} : }

\begin{align}
 f_\omega(z)= \left(1- 4 \pi^2 T^2 z^2 \right)^{- i \frac{\Omega}{2}}(1 - i \Omega h_1(z)-\Omega^2 h_2(z) + i \Omega^3 h_3(z))
\end{align}
where $h_1, h_2$ are known from \eqref{B:h1} and \eqref{B:h2}. The differential equation for
$h_3(z)$ or rather $y_3(z) \equiv h_3^{\prime}(z)$
\begin{align}
y_3^{\prime} (z) + {p_3(z)} y_2(z) = {q_3(z)}
\end{align}
where $p_3(z)=p_1(z)$ and thus integrating factor $I_3(z)=I_1(z)$, as before. 

\begin{align}
y_3(z)&= \frac{c_3}{I_3(z)} +  \frac{1}{I_3(z)} \int^z I_3(x) q_3(x) \mathrm d x \nonumber \\
&= \underbrace{\frac{c_3 z^2}{1-4 \pi^2 T^2 z^2}}_{y_3^h(z)} + \underbrace{\frac{1}{I_3(z)} \int^z I_3(x) q_3(x) \mathrm d x}_{y_3^p(z)}
\end{align}

Now fixing residue of $y_3(z)$ at $z=\frac{1}{2 \pi T}$ to vanish we obtain 
$$c_3 = 0 $$
The functional form of $h_3(z)$ is simple unlike the higher dimensional case.
 \begin{align}\label{B:h2}
  h_3(z) = \pi T z \{Log (1 + 2 \pi T z)\}^2 -\frac{1}{6} \{Log (1 + 2 \pi T z)\}^3
\end{align} \\

Therefore in this perturbative expansion the full solution becomes
\begin{align}
f^R_\omega(z)= \left(1- 4 \pi^2 T^2 z^2 \right)^{- i \frac{\Omega}{2}}\bigg[1 - &i \Omega (\frac{1}{2} ( 4 \pi T z - 2 Log(1+ 2 \pi T z))) -\Omega^2 (\frac{1}{2} (-4 \pi T z + \{Log (1 + 2 \pi T z)\}^2))\nonumber \\ 
+ &i \Omega^3 (\pi T z \{Log (1 + 2 \pi T z)\}^2 -\frac{1}{6} \{Log (1 + 2 \pi T z)\}^3) \bigg]
\end{align}

Evidently in the zero temperature limit (with very small frequency)
\begin{align}\label{B:soln-z}
f^R_\omega(z)\bigg|_{T \to 0} = 1 + \frac{\omega^2 z^2}{2} + i \frac{\omega^3 z^3}{3}
\end{align}

In $r$ co-ordinate 
\begin{align}\label{B:soln-r}
f^R_\omega(r)\bigg|_{T \to 0} = 1 + \frac{\omega^2 L^4}{2 r^2} + i \frac{\omega^3 L^6}{3 r^3}
\end{align}

The retarded Green's function at $T=0$ can be calculated using this solution 
\begin{align}
G^0_R &\equiv \lim_{r \to r_B } T_0(r)f^R_{-\omega}(r) \partial_r f^R_{\omega}(r) \nonumber \\
&=  \lim_{r \to r_B } \frac{1}{2 \pi l_s^2} \frac{r^4}{L^4}\left(-\frac{\omega^2 L^4}{r^3} - i\frac{\omega^3 L^6}{r^4}\right)\nonumber\\
&= -\frac{\mu \omega^2}{2 \pi} - i\frac{\sqrt{\lambda}}{2 \pi} ~\omega^3
\end{align} 

Thus the renormalized Green's function 
\begin{align}
G_R(\omega)\equiv G^0_R + \frac{\mu \omega^2}{2 \pi} =  - i\frac{\sqrt{\lambda}}{2 \pi} ~\omega^3
\end{align}

This matches identically with the leading term in small frequency expansion of \eqref{BTZ GF 0}.


\vspace{2cm}
\centerline{***}

\end{document}